\input{aipcheck}

\documentclass[
    ,final            
  ]
  {aipproc}

\layoutstyle{6x9}

\begin{document}

\title{A Look at Disk--Jet Connections in Stellar--Mass Black Holes}

\classification{Replace this text with PACS numbers}

\author{Jon M. Miller}{
  address={The University of Michigan Department of Astronomy, 500 Church Street, Ann Arbor, MI, 48109-1042, jonmm@umich.edu}
}

\begin{abstract}
Connections between accretion disks and jets in accreting black holes
are anticipated theoretically.  In recent years, potential evidence
for such connections has been emerging, most vividly in the convenient
regime of stellar-mass black holes.  In this contribution, various
lines of evidence for disk--jet connections are briefly examined, from
the standpoint of an observer focused on the role of the disk.  While
many lines of investigation may be promising in the future, obtaining
multi-wavelength lightcurves and correlating jet flux in the radio
band with physical parameters and phenomena tied to the accretion disk
in X-rays may be the most direct.

\end{abstract}

\maketitle


\section{Introduction}

The aim of this contribution is to take a careful look at efforts to
establish connections between accretion disks and jets in accreting
black hole systems.  Research in this area has accelerated in recent
years.  With a host of X-ray and radio observatories now able to bear
on this problem, and as the highly flexible and unique RXTE mission
nears its end, it is prudent to take stock of these advances, to ask
which results are very robust and which need stronger support, and to
try to suggest a course for the near future.

The aims of exploring possible disk--jet connections are many,
including: understanding how and why jets are launched in accreting
black hole systems, understanding how and why jets are quenched in some
circumstances, learning the parameters which govern jet velocity and
power, and understanding the contribution of jets to the radiative
output of accreting black hole systems.

The central observational difficulty is that these issues must be
addressed indirectly in actively accreting systems, since the
innermost accretion flow cannot be imaged in those cases.  Size scales
corresponding to hundreds of gravitational radii (and below) can now
be probed in Sgr A* (e.g., Shen et al.\ 2005) and M87 (Junor, Biretta,
\& Livio 1999), but these black holes appear to accrete at very low
rates, and for the purpose of understanding black hole growth phases
in the early universe, understanding more active accretion phases is
of greater interest.  In this effort, stellar-mass black holes in the
Milky Way represent a convenient regime in which to explore the
processes that may be at work near to supermassive black holes.
Flux--flux correlations, multi-wavelength lightcurves, spectra, and
variability all have a role to play, but at times they can provide
inconsistent or ambiguous clues.  In the sections that follow, we
discuss the extent to which observational evidence has addressed the
interesting questions listed above.  The sources at the center of this
brief examination are the Galactic black holes GRS~1915$+$105, XTE
J1118$+$480, Cygnus X-1, and GX 339$-$4; the neutron star systems Sco
X-1 and Circinus X-1 are also important.

\section{Launching and Powering Jets}

Perhaps the easiest (and maybe the most interesting) question which
observations can address is this: Is black hole spin important in
powering jets?  The potential around a spinning black hole is deeper
than around a non-spinning black hole, and the linear velocity of the
innermost stable circular orbit around a maximally-spinning black hole
approaches $c$, whereas orbits around non-spinning black holes and
neutron stars only approach 0.3--0.4$c$.  If black hole spin is
central to driving jets (for instance, though the Blandford-Znajek
mechanism; Blandford \& Znajek 1977), jet velocities approaching
0.9$c$ should not be seen in accreting neutron star systems.

For a number of years, this separation appeared to hold.  However,
recent observations of the well-known neutron star binary Sco X-1
revealed jets with with a velocity of $v/c \simeq 0.6$ (Fomalont,
Geldzahler, \& Bradshaw 2001).  Even more recent observations of the
neutron star binary Circinus X-1 may have revealed a jet with $v/c
\geq 0.9$ (Fender et al.\ 2004).  This finding may serve to indicate
that black hole spin is likely not the most important ingredient in
driving large-scale radio jets.

Two aspects of the Circinus X-1 result are worth examining: how
certain is the nature of the compact object, and how certain is the
jet velocity implied by observations?  Type-I X-ray bursts were
detected from the direction of Circinus X-1 long ago, which would
certify Circinus X-1 as a neutron star system (Tennant, Fabian, \&
Shafer 1986)).  Some doubts remained, however, regarding whether or
not the bursts were from a background source.  The very recent
detection of two "kilohertz" quasi-periodic oscillations (QPOs) in
Circinus X-1 demands that the X-ray source is indeed a neutron star
(Boutloukos, Wijnands, \& van der Klis 2006).  Thus, all reasonable
doubts about the nature of the central source are resolved.  The
velocity constraints obtained from Circinus X-1 differ from other
sources, in that the velocity is taken from watching standing knots
brighten in sequence (Circinus X-1 has a 16.6-day flux cycle), rather
than tracing a single blob ejected from the central source (as per
"microquasar" systems, and blobs observed in the neutron star binary
Sco X-1).  Only two cycles were followed, and it is possible that the
brightening sequence was not phased correctly, or that fluctuations
downstream do not sequence with the fluctuations of the central
source.

It has been noted that the black hole binaries which have produced the
most extremely relativistic radio jets have the strongest evidence for
spin in X-rays (Fabian \& Miller 2002).  It has also been noted that
the most relativistic jets and ejection events have been seen in
long-period black hole binaries (Garcia et al.\ 2003), which may
already hint that if black hole spin is involved in powering jets, it
may not be the only important parameter.  A careful, high-cadence
monitoring campaign on Circinus X-1 will be central to making a
rigorous statement about the role of black hole spin in powering jets.

\section{Jet Quenching Mechanisms}

It has been shown that in both stellar-mass and supermassive black
holes, radio and X-ray fluxes are correlated over many orders of
magnitude at low fractions of the Eddington mass accretion rate (below
0.01--0.1 $L_{Edd}$; see, e.g., Gallo, Fender, \& Pooley 2003;
Merloni, Heinz, \& Di Matteo 2003; Falcke, Kording, \& Markoff 2004;
and Merloni et al.\ 2006).  At the higher mass accretion rates
observed in Seyfert AGN, radio flux appears to be greatly diminished
(though weak radio activity and even resolved compact jets are seen in
some cases).  At high fractions of the Eddington limit and in soft,
disk--dominated phases, radio emission in stellar-mass black hole
systems and X-ray binaries appears to be quenched (Gallo, Fender, \&
Pooley 2003; Maccarone, Gallo, \& Fender 2003).

At present, the nature of this quenching is not well understood.  A
simple expectation, based partly on the predictions of advection
dominated accretion flow models (ADAFs) might be that a filled inner
disk in high luminosity states may be involved in quenching jets
(Fender, Belloni, \& Gallo 2004).  New work appears to rule-out this
possibility: accretion disks appear to remain close to the black hole
both in high luminosity and low luminosity phases alike (for
$L_{X}/L_{Edd} = 0.001$, at least; Miller et al.\ 2006; Miller,
Homan, \& Miniutti 2006).  It is possible that the mass accretion rate
through the disk, more than the radius of the inner disk, may act to
control jet production.  It is also possible that a parameter like the
ratio of energy dissipated in an electron-dominated corona to energy
dissipated in the disk, may act to control jet production; this
parameter would be at a minimum in high luminosity, disk--dominated
states.

It has previously been suggested that jet production may be tied to
the formation of poloidal magnetic fields (Blandford \& Payne 1982;
for a recent example, see Livio, Pringle, \& King 2003).  This is a
particularly interesting possibility.  The absence of a poloidal field
in soft, disk--dominated phases, then, may inhibit jet production.
This possibility is qualitatively consistent with the observation of a
magnetically-driven wind in a soft disk--dominated state of the
stellar mass black hole GRO~J1655$-$40 (Miller et al. 2006b).  In that
source, the magnetic wind is not likely to be magnetocentrifugal (due
to poloidal fields), due to its high mass outflow rate (Proga 2003).
It is more likely driven by magnetic pressure from tangled fields in
the disk.  It has also been noted in the stellar-mass black hole
GX~339$-$4 and black hole candidate H~1743$-$322 that winds appear to
be active in soft, disk-dominated states, and to be absent in hard
states (Miller et al. 2004, 2004b, 2006c).  Moreover, the absence of wind
absorption in hard phases in H~1743$-$322 and GX~339$-$4 cannot be due
only to ionization; either the geometry must change or the wind must
be quenched (Miller et al. 2004b, 2006c).

Thus, observations may support a picture in which jets are launched
when poloidal fields can be maintained (e.g., when the ratio of energy
dissipated in hard X-rays to that dissipated in the disk is above some
threshold), and quenched when poloidal fields give way to toroidal
fields, energy dissipation dominated by viscosity internal to the
disk, and disk winds.

\section{Broad-band Spectral Distributions}

One important means of understanding the role of jets in
accretion-powered systems to observe and monitor the broad-band
spectral energy distribution, from radio up to hard X-rays.  Radio
imaging clearly reveals continuous jets in some cases (e.g., Cygnus
X-1; Stirling et al.\ 2001), and rapidly moving blobs in others (e.g.,
GRS~1915$+$105; Fender et al.\ 1999), so it is clear that jets produce
radio synchrotron emission downstream.  The harder question to answer
is: How much do jets contribute to higher frequency emission closer to
the compact object?

In recent years, considerable effort has been devoted to trying to
model the broad-band spectral energy of black holes, with and without
jets, in an effort to understand their role in the overall radiative
output observed from black holes.  A critical test case for these
efforts is the black hole transient XTE~J1118$+$480, which yielded
excellent UV data due to its high galactic latitude (and therefore its
low extinction).  Both jet-dominated emission models (Markoff, Falcke,
\& Fender 2001), and broad-band models assuming an advective inner
flow (McClintock et al.\ 2001, Esin et al.\ 2001), appear to give
broadly reasonable descriptions of the data.  The jet-dominated
emission models fit the IR data slightly better than the more
traditional broad-band model with an advective inner flow.  However,
more such broad-band spectral energy distributions are required to
better differentiate these families of models.

Another regime in which jet emission might be differentiated from more
traditional models with a Comptonizing corona or inner advective flow,
is in the hard X-ray band (somewhat arbitrarily, above 100~keV).  Jet
emission -- if primarily due to synchrotron emission -- should not
turn over abruptly in the hard X-ray band.  Emission from a
Comptonizing corona, however, should turn over at high energy, due to
a decrease in the scattering cross section at energies approaching the
rest mass of the electron.  Some spectra appear to show that such a
turn-over exists (e.g. Zdziarski \& Gierlinski 2004), which may
suggest that jet emission does not extend up to the hard X-ray band.
Background emission can be very high in hard X-ray bands, however, and
instruments which could obtain such spectra (e.g., OSSE aboard {\it
CGRO}) typically have very large fields of view.  Such results must be
viewed with these cautions in mind.  As the {\it INTEGRAL} mission
continues, and the detector performance and backgrounds are better
understood, it may be possible to obtain more robust spectra above
100~keV.  The {\it Suzaku} HXD may also offer sensitive hard X-ray
spectra in the near future.

In fact, the situation is even more complex than described above.  It
is expected that hard X-ray emission should cause a reaction in the
accretion disk known as disk reflection (e.g., George \& Fabian 1991,
Nayakshin \& Kallman 2001, Ross \& Fabian 2005) .  The most prominent
results of this process are an Fe~K emission line, and a "reflection
hump" (actually due to Compton back-scattering in the disk)
which peaks at approximately 30 keV, where it appears as subtle
curvature in a positive flux sense.  Disk reflection makes it
particularly difficult to detect curvature which might be due to
thermal Comptonization in a corona apart from that due to disk
reprocessing -- if a break in the power-law exists near to 30~keV, it
may be lost due to extra curvature in this region.  New jet models
place additional emphasis on synchrotron self-Comptonization, not just
direct synchrotron emission, which may further complicate efforts to
discern jet-based emission from standard coronal emission in accreting
black holes.

\section{Multi-wavelength Lightcurves}

Multi-wavelength lightcurves may offer a more incisive means of
parsing the roles of disks, jets, and/or hard X-ray coronae in
creating the broad-band spectra observed from accreting black holes.
In an IR, optical, UV, and X-ray variability study of the flux
observed in XTE J1118$+$480, for instance, Hynes et al.\ (2003)
conclude that the dominant variability mechanism is not simple disk
reprocessing, and suggest that the flux may instead have a synchrotron
origin.  This finding is consistent with model for jet-based emission
(Markoff, Falcke, \& Fender 2001), and likely inconsistent with more
standard models.

Similar findings have recently been reported based on a
multi-wavelength study of GX~339$-$4.  In spectrally hard X-ray
states, the optical and IR flux are observed to trace the X-ray flux
very closely, in the same manner that X-ray flux and radio jet flux
are correlated, without a measurable delay (Homan et al.\ 2005).  This
suggests that a large fraction of the optical and IR flux may arise
via non-thermal processes in the jet, rather than due to reprocessing
of X-ray flux in the outer disk.

Thus, whereas it can be difficult to distinguish jet-based emission
models from more standard models based on broad-band spectroscopy, the
situation appears more clear when multi-wavelength lightcurves are
examined in detail.  While these findings may demonstrate that jets
are very important, in another regard their prominence is a hindrance:
IR, optical, and UV emission which was formerly associated with the
disk, may actually be due to the jet.  If it is not possible to
clearly distinguish jet emission from disk emission, learning the
nature of disk--jet coupling in black holes will be very difficult.
In this regard, searching for correlations between jet-based radio
flux and X-ray phenomena tied to the disk (thermal continuum emission,
Fe emission lines, QPOs) may be the best way forward.

\section{X-ray--Radio Connections in Black Holes}

As noted above, a number of recent studies have clearly demonstrated a
link between hard X-ray flux and radio flux in accreting black hole
systems.  Indeed, flux correlations hold in both stellar-mass black
holes and supermassive black holes, and extend over several orders of
magnitude.  The correlations hold in low flux states, which are
generally dominated by hard X-ray emission.

These flux--flux correlations are the foundation of many
investigations into disk--jet coupling in black holes.  But there is a
potential problem: if the bulk of the hard X-ray emission seen in
black holes arises in the jet, and is not closely tied to the disk,
the flux--flux correlations seen may not be disk--jet connections, but
jet--jet connections.  The missing link in studies of multi-wavelength
fluxes, is how aspects of the X-ray flux which may be reliably
assigned to the disk (not the corona or jet) vary with optical, IR,
and radio flux.  This may be the only clear means of understanding how
the accretion inflow and outflow are related.

\subsection{A Careful Look at the Disk-Jet Connection in GRS~1915$+$105}

The Galactic microquasar GRS~1915$+$105 is well-known to both X-ray
and radio observers as a bright, variable accretor.  GRS~1915$+$105 may
be most notable for the radio-emitting blobs it sometimes launches at
high velocities ($v/c \geq 0.9$).  As the source is bright and has
been active for years, it is a natural target in which to explore
disk--jet connections.

Perhaps the most exciting evidence for a disk--jet connection in GRS
1915$+$105, if not in accreting black holes in general, was presented
by Belloni et al.\ (1997).  In a particular observation made with {\it
RXTE}, dramatic swings in the source intensity and hardness over
$\sim$1000~s spans can be explained in terms of rapid changes in the
inner accretion disk radius.  This behavior is highly suggestive of a
disk which is periodically being "emptied".  Pooley \& Fender (1997)
discovered a link between this apparent behavior and radio flaring,
indicating that the disk might be periodically ejected in a jet.

In order to test this connection, we analyzed the same {\it RXTE}
dataset considered in Belloni et al.\ (1997).  Using simple count-rate
selections, we made average spectra from the high ("filled-disk") and
low ("emptied disk") phases in the 2.8-20.0 keV band.  All layers of
all active PCUs were combined to create the spectra.  Responses were
generated using the standard "pcarsp" tool.  The spectra were fit
jointly using XSPEC version 11.  We fit the spectra with a simple model
consisting of Galactic absorption, an accretion disk (using the
simple "diskbb" model), and Comptonization (using the "CompTT" model).
Within this model, the seed temperature in the Comptonization model
was fixed to the disk temperature.  


With this simple spectral model, we find that the apparent ejection
episodes can be instead described in terms of a disk which remains at
the innermost stable circular orbit, and a variable corona.  A disk
with a color temperature of $kT = 1.29(1)$~keV and an apparent radius
of $2~r_g$ (where $r_g = GM/c^2$, and assuming $d=11$~kpc, an
inclination of 70 degrees, and a mass of 14 solar masses) with a $kT_e
= 50$~keV Comptonizing corona changing between $\tau < 0.05$ in high
phases and $\tau = 0.29(1)$ in low phases gives an acceptable fit to
the data ($\chi^{2}/\nu = 104/86$).  These results suggest that the
corona, not the disk, might be periodically ejected in GRS~1915$+$105.
A similar conclusion was reached by Vadawale et al.\ (2003) using a
more complex spectral model.

Our spectral results from GRS~1915$+$105 are not unique; a variable
disk as per Belloni et al.\ (1997) can also describe the variability
acceptably.  Our results only serve to indicate that present evidence
for a disk--jet connection in GRS~1915$+$105 is not definitive.  This
case may serve as an example of the various observational challenges
in clearly demonstrating a disk--jet connection.  {\it RXTE} has a
lower energy bound of $\sim$3~keV and low energy resolution, which can
complicate studies of accretion disks.  Moreover, GRS~1915$+$105 is
viewed through a very high Galactic column density, which further
complicates studies of the disk.

\section{Jets and X-ray Timing}

X-ray power spectral components are believed to be tied to the disk,
and can therefore be an important tool for testing disk--jet
connections.  There are subtle complications, however.  For instance,
the rms power in such components is positively correlated with X-ray
hardness (e.g., Belloni et al.\ 2005).  Thus, the energy dependence of
timing components is not disk-like, but corona-like.  Within the
context of this article, then, where the necessity of establishing jet
correlations with true disk parameters is emphasized, it is worth
briefly discussing why it is believed that QPOs are disk phenomena.

It is likely that the disk supplies frequencies which may be more
vividly expressed in a corona.  The best evidence for this may be come
from the behavior of QPOs observed in Sco X-1: the exact modulation
mechanism of kHz QPOs in neutron stars is not known, but the
frequencies of kHz QPOs make it very likely that these QPOs reflect
the orbital frequency of the inner disk and an interaction with the
stellar surface (e.g. Barret, Olive, \& Miller 2006).  Yu, van der
Klis, \& Jonker (2001) showed that the properties of these kHz QPOs
vary with the phase of slower oscillations.  This finding suggests
that low-frequency QPOs, while not orbital frequencies, are still disk
frequencies, perhaps due to precession or a global disk oscillation.

Similarly, the frequencies of the fastest QPOs observed in stellar-mass
black hole systems is very close to Keplerian orbital frequencies
expected at the innermost stable circular orbit.  Once again, the
nature of lower frequency QPOs is less clear, but indirect evidence
suggests they are also tied to the disk.  For instance, Miller \& Homan
(2005) showed that the properties of broad Fe lines -- believed to
originate in the inner disk -- depend on the phase of strong
low-frequency QPOs in GRS~1915$+$105.  As with the Yu result, this
suggests that low-frequency QPOs may be a global oscillation or a
precession frequency.

Migliari, Fender, \& van der Klis (2005) conducted the first study of
X-ray timing components and radio jet flux in the black hole
GX~339$-$4, and a number of neutron star binaries.  The exciting
result of this study is that radio luminosity appears to be positively
correlated with one Lorentzian component and a frequency break, and
anti-correlated with the rms power in that specific Lorentzian, in
GX~339$-$4 (individually; the same holds true for the collection of
less--observed neutron star sources when considered together).  

This correlation is very suggestive of a link between frequencies in the accretion
disk and power in the radio jet.  Two caveats must be noted:
First, the study correlates the properties of Lorentzian power
components and frequency breaks in the X-ray flux with the radio flux;
while it is likely that broad Lorentzian components are related to
QPOs, they are not properly QPOs (the term "QPO" is typically reserved
for Lorentzian components for which $\nu/FWHM \geq 2$).  Second, the
Lorentzian with properties found to correlate with radio luminosity is
just one of a few Lorentzians required to fit the X-ray power spectra.

The main advantage and disadvantage of correlations between radio
parameters and X-ray timing parameters is that the timing components
are phenomenological.  The timing components do not suffer from the
same degree of model degeneracy that complicates disk--jet connections
in GRS~1915$-$105 using continuum spectra.  However, the fact that the
precise origin of power spectral components in accreting systems is
still not understood prevents these correlations from providing
clearer insights.

\section{Future Directions}

As discussed above, multi-wavelength lightcurves have clearly shown
that jets may be very important in accreting black hole systems.  To
reveal any coupling between the accretion disk and jet, however, more
detailed work is required.  It is also clear that we must move away
from simply exploring radio correlations with broad-band X-ray flux,
or hard X-ray flux.

In the case of stellar-mass black holes, the accretion disk can be
revealed in three ways: through thermal blackbody emission, through
(broad) Fe emission lines, and through X-ray timing features.  Of
these, our brief analysis of GRS~1915$+$105 suggests that thermal
continuum emission may be the most ambiguous of the potential disk
diagnostics.  The accretion disk reflection spectrum -- most vividly
seen in Fe emission lines (which are relatively independent of the
continuum, see Miller et al. 2006) -- is a much better diagnostic.
X-ray timing components may also be good disk diagnostics,
particularly if new studies can employ QPOs rather than broader
Lorentzian features.

For significant progress, it is likely that as many as 100 points from
which radio flux, radio spectral index, Fe line properties, and X-ray
timing properties can me measured simultaneously, over the course of
an outburst.  This is a high standard, to be sure, but to really
understand disk--jet connections, we must be able to probe beyond
simple mass accretion rate variations.  A large number of points will
provide the ability to test whether a given correlation is
statistically better than broad-band flux--flux correlations (taking
flux as an imperfect indicator of mass accretion rate).

These studies will require more total exposures, and more frequent
monitoring of outbursts with missions like {\it Chandra}, {\it
XMM-Newton}, {\it Suzaku}, and {\it Swift}, and more focused studies
with RXTE in its final years of operation.  While RXTE can constrain
the basic parameters of Fe lines, missions with lower background and
higher energy resolution are better suited.  {\it RXTE} is, however,
ideally suited to provide the timing measurements required.

\section{Acknowledgements}

I thank Jeroen Homan and Simone Migliari for helpful discussions.  I
thank Philip Hughes, Joel Bregman, and the University of Michigan for
arranging an excellent conference.


\begin{thebibliography}{9}


\bibitem[]{} Barret, D., Olive, J., \& Miller, M., 2006, MNRAS, in press

\bibitem[]{} Belloni, T., et al., 2005, A\&A, 440, 207

\bibitem[]{} Blandford, R., \& Payne, D., 1982, MNRAS, 199, 883

\bibitem[]{} Blandford, R., \& Znajek, R., 1977, MNRAS, 179, 433

\bibitem[]{} Belloni, T., Mendez, M., King, A. R., van der Klis, M.,
\& van Paradijs, J., 1997, ApJ, 479, L145

\bibitem[]{} Boutloukos, S., Wijnands, R., \& van der Klis, M., 2006, ATEL 695

\bibitem[]{} Esin, A., et al., 2001, ApJ, 555, 483

\bibitem[]{} Fabian, A., \& Miller, J., 2002, Science, 297, 947

\bibitem[]{} Falcke, H., Kording, E., \& Markoff, S., 2004, A\& A, 414, 895

\bibitem[]{} Fender, R., et al., 1999, MNRAS, 304, 865

\bibitem[]{} Fender, R., et al., 2004, Nature, 427, 222

\bibitem[]{} Fender, R., Belloni, T., \& Gallo, E., 2004, MNRAS, 335, 1105

\bibitem[]{} Fomalont, E., Geldzahler, B., \& Bradshaw, C., 2001, ApJ, 553, L27

\bibitem[]{} Gallo, E., Fender, R., \& Pooley, G., 2003, MNRAS, 344, 60

\bibitem[]{} Garcia, M., Miller, J., McClintock, J., King, A., \& Orosz, J., 2003,
ApJ, 591, 388

\bibitem[]{} George, I., \& Fabian, A., 1991, MNRAS, 249, 352

\bibitem[]{} Homan, J., Buxton, M., Markoff, S., Bailyn, C., Nespoli, e., \&
Belloni, T., 2005, ApJ, 624, 295

\bibitem[]{} Hynes, R., et al., 2003, MNRAS, 345, 292

\bibitem[]{} Junor, W., Biretta, J., \& Livio, M., 1999, Nature, 401, 891

\bibitem[]{} Livio, M., Pringle, J., \& King, A., 2003, ApJ, 593, 184

\bibitem[]{} Maccarone, T., Gallo, E., \& Fender, R., 2003, MNRAS, 345, L19

\bibitem[]{} Markoff, S., Falcke, H., \& Fender, R., 2001, A\&A, 372, L25

\bibitem[]{} McClintock, J., et al., 2001, ApJ, 555, 477

\bibitem[]{} Migliari, S., Fender, R., \& van der Klis, M., 2005, MNRAS, 363, 112

\bibitem[]{} Miller, J. M., et al., 2004, ApJ, 601, 450

\bibitem[]{} Miller, J. M., et al., 2004b, ATEL 221

\bibitem[]{} Miller, J., \& Homan, J., 2005, ApJ, 618, L107

\bibitem[]{} Miller, J. M., et al., 2006, ApJ, subm., astro-ph/0602633

\bibitem[]{} Miller, J. M., et al., 2006b, Nature, 441, 953

\bibitem[]{} Miller, J. M., et al., 2006c, ApJ, in press

\bibitem[]{} Miller, J. M., Homan, J., \& Miniutti, G., 2006, ApJ, subm.,
astro-ph/0605190

\bibitem[]{} Merloni, A., Heinz, S., \& Di Matteo, T., 2003, MNRAS, 345, 1057

\bibitem[]{} Merloni, A., Kording, E., Heinz, S., Markoff, S., Di Matteo, T., \&
Falcke, H., 2006, NewA, 11, 567

\bibitem[]{} Nayakshin, S., \& Kallman, T., 2001, ApJ, 546, 406

\bibitem[]{} Pooley, G. G., \& Fender, R. P., 1997, MNRAS, 292, 925

\bibitem[]{} Proga, D., 2003, ApJ, 585, 406

\bibitem[]{} Ross, R., \& Fabin, A., 2005, MNRAS, 358, 211

\bibitem[]{} Shen, Z., Lo, K., Liang, M., Ho, P., \& Zhao, J., 2005, Nature, 438, 62

\bibitem[]{} Stirling, A., Spencer, R., de la Force, C., Garrett, M., Fender, R.,
\& Ogley, R., 2001, MNRAS, 327, 1273

\bibitem[]{} Tennant, A., Fabian, A., \& Shafer, R., 1986, MNRAS, 221, 27

\bibitem[]{} Vadawale, S. V., Rao, A. R., Naik, S., Yadav, J. S.,
Ishwara-Chandra, C. H., Pramesh Rao, A., \& Pooley, G. G., 2003, ApJ,
597, 1023

\bibitem[]{} Yu, W., van der Klis, M., \& Jonker, P., 2001, ApJ, 559, L29

\bibitem[]{} Zdziarski, A., \& Gierlinski, M., 2004, PThPS, 155, 99

\end{thebibliography}
\end{document}